# An algorithm for finding the Independence Number of a graph

O.Kettani

email :o_ket1@yahoo.fr

**Abstract**

In this paper, we prove that for every connected graph G, there exists a split graph G' with the same independence number and the same order . Then we propose a first algorithm for finding this graph, given the degree sequence of the input graph G. Further, we propose a second algorithm for finding the independence number of G, given the adjacency matrix of G.

**Introduction**

Let G=(V,E) be a connected graph on n=│V│ vertices and m=│E│ edges.
A subset X of V is called independent if its vertices are mutually non-adjacent. The independence number α(G) is the largest cardinality among all independent sets of G.
A split graph is a graph in which the vertices set can be partitioned into a clique and an independent set.
The problem of finding the independence number α of a graph G is know to be NP- hard[1].
In this paper, we prove that for every connected graph G, there exists a split graph G' with the same independence number and the same order . Then we propose an algorithm for finding this graph, given the degree sequence of the input graph G. Further, we propose a second algorithm for finding the independence number of G, given the adjacency matrix of G.
First we will prove the following proposition :

**Proposition 1** :
For every connected graph G, there exists a split graph G' with the same independence number α and the same order n. The degree sequence of G' contains (n- α) times the degree (n-1) and α times the degree (n-α).

**Proof**:

Let G=(V,E) be a connected graph on n=|V| vertices and independence number α.
Then ∃I ⊂V, I independent set, with α = |I| and ∃VC ⊂V (VC a vertex cover) such
V=I∪VC and I∩VC= ∅.
Considere the graph G' constructed from G by the following rules:

1) For every vertex v in I, and for every vertex w in VC non-adjacent to v create an edge between v and w.
2) If |VC| ≥2 then for every non-adjacent vertices v and w in VC, create an edge between v and w.

Then clearly the resulting graph G' is such that V=V' and I =I' and VC=VC'
and I'∩VC'= ∅.
Since the subgraph induced by VC' is a clique then G' is a split graph and,

$|VC|=n-\alpha$ and $v\in VC \Rightarrow d(v)= \alpha+(n-\alpha)-1=n-1$

$|I|= \alpha$ and $v\in I \Rightarrow d(v)= n-\alpha$

Thus, the degree sequence of G' contains (n- α) times the degree (n-1) and α times the degree (n-α).
Let $G_{\alpha,n-\alpha}$ denote the split graph G'

**Example** : n=6 and G is the following graph :

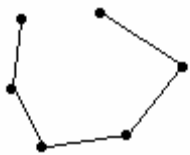

the resulting graph G' is $G_{3,3}$:

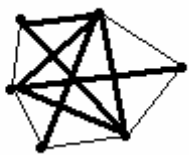

For n=3, $G_{2,1}$ is the the following graph :

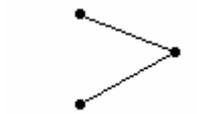

And is $G_{1,2}$ is the triangle $K_3$

For n=4, $G_{3,1}$ is the the following graph :

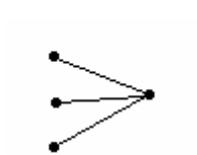

$G_{2,2}$

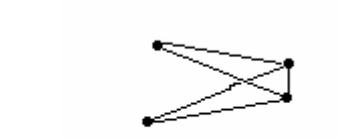

$G_{1,4}$

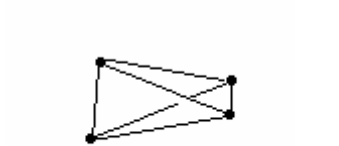

For n=5, $G_{4,1}$ is the the following graph :

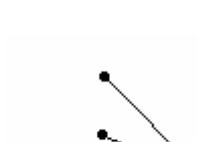

$G_{3,2}$

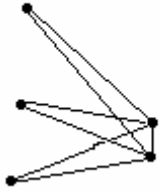

$G_{2,3}$

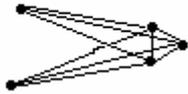

$G_{1,4}$ is $K_5$

For n=6, $G_{5,1}$ is

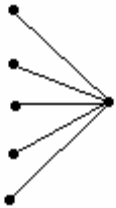

$G_{4,2}$ is

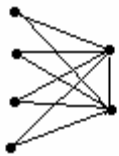

$G_{3,3}$ is

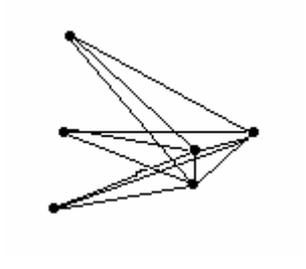

$G_{2,4}$ is

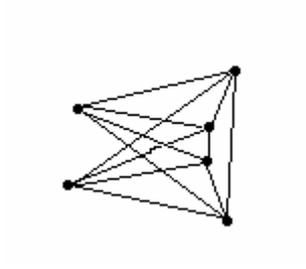

And $G_{1,5}$ is $K_6$

and so on…

**Definition** 1 :

Given two degree sequence $S = (d_1 \geq d_2 \geq ... \geq d_n)$ and $S' = (d'_1 \geq d'_2 \geq ... \geq d'_n)$

We denote $S' \geq S$ iff $d'_i \geq d_i$ for all i.

And if $S' \geq S$, we define $S'-S = (d'_1 - d_1, ..., d'_n - d_n)$

**Definition** 2 :

A degree sequence S is said realizable (or graphic) iff it represents a simple graph.

**Definition** 3 :

Let $G=(V,E)$ be a connected graph and $G'=(V,E')$ be an other connected graph such $E \subset E'$.

Then we define the difference graph G'', by $G''=(V,E'-E)$.

Then $S(G'') = S(G') - S(G)$.

Before establishing the main result of the present paper, we will use the following deterministic procedure HH (S) due to Havel [3] and Hakimi [4], which can be used to decide whether a degree sequence is or is not realizable (i.e.,represents a simple graph or not). It is based on the Erdös-Gallai theorem [2] which states that given a degree sequence $S = (d_1 \geq d_2 \geq ... \geq d_n)$, S is realizable if and only if the following equation holds :

$$\sum_{i=1}^{k} d_i \leq k(k-1) + \sum_{i=k+1}^{n} \min(k, d_i)$$

The Havel-Hakimi procedure tests the degree sequence S as follows.

procedure HH(S)

Input $S = (d_1 \geq d_2 \geq ... \geq d_n)$,

1 If there exists an integer d in S such that $d > n-1$ then halt and output false. That is, we cannot have a vertex that is adjacent to more than $n - 1$ other vertices.

2 If there are an odd number of odd numbers in S halt and and output false. That is, there must be an even number of vertices of odd degree.

3 If the sequence contains a negative number then halt and and output false.

4 If the sequence is all zeros then halt and output True

5 Reorder S such that it is nonincreasing.

6 Delete the first term $d_1$ from S and subtract one from the next $d_1$ terms to form a new sequence.

Go to step 3 .

**Proposition 2 :**

Let G=(V,E) be a connected graph, then there exists a unique split graph $G_{\alpha,n-\alpha}$ such that $S(G_{\alpha,n-\alpha}) \geq S(G)$ and the degree sequence $S(G_{\alpha,n-\alpha}) - S(G)$ is realizable.

**Proof** :

The existence of the split graph $G_{\alpha,n-\alpha}$ is proved by proposition 1.

Since $E(G) \subset E(G_{\alpha,n-\alpha})$ then $S(G_{\alpha,n-\alpha}) \geq S(G)$ and $S(G_{\alpha,n-\alpha}) - S(G)$ is realizable because it represents the difference graph $G''=(V, E(G_{\alpha,n-\alpha}) - E(G))$.

Suppose on the contrary that there exists another split graph $G_{\alpha',n-\alpha'}$ such that $S(G_{\alpha',n-\alpha'}) \geq S(G)$ and the degree sequence $S(G_{\alpha',n-\alpha'}) - S(G)$ is realizable.

Then by lemma 1, $\alpha = \alpha(G) = \alpha(G_{\alpha',n-\alpha'}) = \alpha'$ : a contradiction. Consequently, there exists a unique split graph $G_{\alpha,n-\alpha}$ such that $S(G_{\alpha,n-\alpha}) \geq S(G)$ and the degree sequence $S(G_{\alpha,n-\alpha}) - S(G)$ is realizable.

Now, we will prove the first main result of this paper :

**Proposition 3** :

Given a connected graph $G=(V,E)$ on $n=|V|$ vertices and m edges with degree sequence $S(G) = (d_1 \geq d_2 \geq ... \geq d_n)$, then its independence number $\alpha$ could be computed by the following algorithm :

Input : $S(G) = (d_1 \geq d_2 \geq ... \geq d_n)$.

1 $\alpha \leftarrow 1$

2 Repeat

    if $S((G_{\alpha,n-\alpha})) \geq S(G)$ and $HH(S(G_{\alpha,n-\alpha}) - S(G))$ then output($\alpha$) halt

                                                         Else $\alpha \leftarrow \alpha+1$ goto step 2

End.

**Proof** :

By using proposition 1 and 2, the algorithm finds the unique « closest » split graph $G_{\alpha,n-\alpha}$ and then its independence number $\alpha$ which is also the independence number of G.

**Example** : n=6 and G is the following graph :

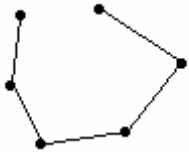

$S(G) = (2,2,2,2,1,1)$.

$S((G_{1,5})) = (5,5,5,5,5,5)$

$S((G_{1,5})) \geq S(G)$ but $HH(S(G_{\alpha,n-\alpha}) - S(G)) = HH((3,3,3,3,4,4))$ is false.

$S((G_{2,4})) = (5,5,5,5,4,4)$ but $HH(S(G_{\alpha,n-\alpha}) - S(G)) = HH((3,3,3,3,3,3))$ is false.

$S((G_{3,3})) = (5,5,5,3,3,3)$ and $HH(S(G_{\alpha,n-\alpha}) - S(G)) = HH((3,3,3,1,2,2))$ is true.

Then the algorithm outputs $\alpha = 3$

Let n be an integer ≥3 . Let θ be the nxn null matrix.

For k such 1≤k≤n-1, we define the triangular matrices $T_k$ by :

$(T_{k+1})_{i,j}=1$  if   1≤i≤k and 2≤j≤k+1 and i<j

$(T_{k+1})_{i,j}=0$  otherwise.

$$T_{k+1} = \begin{pmatrix} 0 & 1 & \cdots & 1 & 0 & \cdots & 0 \\ & 0 & \ddots & \vdots & \vdots & & \vdots \\ & & \ddots & 1 & 0 & & \\ 0 & \cdots & & 0 & 0 & \cdots & 0 \\ & & & & 0 & & \vdots \\ \vdots & & & & & \ddots & 0 \\ 0 & & \cdots & & & & 0 \end{pmatrix} \leftarrow k+1$$

with column k+1 indicated.

$$T_2 = \begin{pmatrix} 010 & \cdots & 0 \\ 000 & & \\ & & \vdots \\ \vdots & & \\ 0 & \cdots & 0 \end{pmatrix} \quad T_3 = \begin{pmatrix} 0110 & \cdots & 0 \\ 0010 & & \\ 0000 & & \\ & & \vdots \\ \vdots & & \\ 0 & \cdots & 0 \end{pmatrix}$$

$$T_4 = \begin{pmatrix} 01110 & \cdots & 0 \\ 00110 & & \\ 00010 & & \\ 00000 & & \\ & & \vdots \\ \vdots & & \\ 0 & \cdots & 0 \end{pmatrix} \cdots \quad T_n = \begin{pmatrix} 011 & \cdots & 1 \\ 001 & & \\ & \ddots & \vdots \\ & & 1 \\ 0 & \cdots & 0 \end{pmatrix}$$

**Proposition 4 :**

For k such $1 \leq k \leq n-1$, $(T_{k+1})^{k+1} = \theta$

And $(T_{k+1})^h \neq \theta$ if h is such $1 \leq h \leq k$.

**Proof** :

For k such $1 \leq k \leq n-1$, let $C_{k+1}$ denote the matrice defined by:

$(C_{k+1})_{i,j} = 1$ if $j = k+1$ and $1 \leq i \leq k$

$(C_{k+1})_{i,j} = 0$ otherwise.

$$C_{k+1} = \begin{pmatrix} 0 & \cdots & 0 & \overset{k+1}{\underset{\downarrow}{1}} & 0 & \cdots & 0 \\ \vdots & & \vdots & \vdots & & & \vdots \\ \vdots & & \vdots & 1 & 0 & & \vdots \\ 0 & \cdots & 0 & 0 & 0 & \cdots & 0 \\ \vdots & & & 0 & \ddots & & \vdots \\ \vdots & & & & \ddots & \ddots & 0 \\ 0 & & \cdots & & & & 0 \end{pmatrix} \leftarrow k+1$$

For k such $1 \leq k \leq n-1$, we define the set of nxn matrices $\mathbf{M}(k)$ by :

$M \in \mathbf{M}(k)$ iff:

$(M)_{i,j}$ integer $> 0$ for $j = k+1$ and $1 \leq i \leq k$

$(M)_{i,j} = 0$ otherwise..

Then :

For k such $2 \leq k \leq n-1$, we have:

$T_{k+1} = T_k + C_{k+1}$

$(T_{k+1})^2 = (T_k)^2 + T_k C_{k+1} + C_{k+1} T_k + (C_{k+1})^2$

since $(C_{k+1})^2=\theta$ and $C_{k+1}T_k=\theta$ and $T_kC_{k+1}\in M(k-1)$

$$T_kC_{k+1}=\begin{pmatrix} 0 & \cdots & 0 & \overset{k-1}{0} & 0 & \cdots & 0 \\ \vdots & & & k-2 & & & \vdots \\ \vdots & & & \ddots & & & \vdots \\ \vdots & & & & 1 & 0 & \vdots \\ 0 & \cdots & & 0 & 0 & 0 & \cdots & 0 \\ \vdots & & & & & 0 & \ddots & \vdots \\ \vdots & & & & & & \ddots & 0 \\ 0 & & \cdots & & & & & 0 \end{pmatrix} \leftarrow k-1$$

with $k+1$ indicated above.

then

$(T_{k+1})^2=(T_k)^2+ T_kC_{k+1}$ and $T_kC_{k+1}\in M(k-1)$

$(T_{k+1})^3=(T_k)^3+ T_k(T_kC_{k+1})$ since $T_kC_{k+1}\in M(k-1)\Rightarrow T_k(T_kC_{k+1}) \in M(k-2)$

then :

$(T_{k+1})^3=(T_k)^3+ (T_k)^2C_{k+1}$ and $(T_k)^2C_{k+1}\in M(k-2)$

Suppose by induction on $p<k-1$ that:

$(T_{k+1})^p=(T_k)^p+ (T_k)^{p-1}C_{k+1}$ and $(T_k)^{p-1}C_{k+1}\in M(k-p+1)$

then

$(T_{k+1})^{p+1}=(T_k)^{p+1}+ T_k(T_k)^{p-1} C_{k+1}$ and $T_k(T_k)^{p-1}C_{k+1}\in M(k-p)$

Since the induction is true for $p=1$

Then for $p$ such $k>p\geq 1$ we have :

$(T_{k+1})^{p+1}=(T_k)^{p+1}+(T_k)^p C_{k+1}$ and $(T_k)^p C_{k+1}\in M(k-p)$

Then for $p=k-1$ we have :

$(T_{k+1})^k=(T_k)^k+(T_k)^{k-1}C_{k+1}$ and $(T_k)^{k-1}C_{k+1}\in M(1)$

since $(T_k)^{k-1}C_{k+1}\in M(1)\Rightarrow T_k(T_k)^{k-1}C_{k+1}=\theta$

Then

$(T_{k+1})^{k+1}=(T_k)^{k+1}+(T_k)^k C_{k+1}$

Then

$(T_{k+1})^{k+1}=(T_k)^{k+1}$ \hfill (1)

$(T_2)^2=\theta$

Suppose by induction on k that: $(T_k)^k = \theta$ then by equation (1) we have:

$(T_{k+1})^{k+1} = \theta$

On the other hand, for p such $1 \leq p \leq k$:

$(T_{k+1})^p = (T_k)^p + (T_k)^{p-1} C_{k+1}$ and $(T_k)^{p-1} C_{k+1} \in M(k-p+1)$

since $1 \leq p \leq k \Rightarrow k-p+1 \geq 1 \Rightarrow (T_k)^{p-1} C_{k+1} \neq \theta$ and $(T_{k+1})^p \neq \theta$ for p such $1 \leq p \leq k$.

Then the induction is proved.

For k such $1 \leq k \leq n-2$, we define the matrices $B'_{k+1}$ by :

$(B'_{k+1})_{i,j} = 1$ if $1 \leq i \leq k$ and $k+2 \leq j \leq n$

$(B'_{k+1})_{i,j} = 0$ otherwise.

$$B_{k+1} = \begin{pmatrix} 0 & \cdots & & 0 & 1 & \cdots & 1 \\ \vdots & \ddots & & & \vdots & & \vdots \\ & & & & 0 & 1 & \cdots & 1 \\ 0 & \cdots & & 0 & 0 & \cdots & 0 \\ & & & & 0 & \ddots & \vdots \\ \vdots & & & & & \ddots & 0 \\ 0 & \cdots & & & & & 0 \end{pmatrix} \begin{matrix} \\ \\ \leftarrow k \\ \\ \\ \\ \end{matrix}$$

with $k+1$ indicated above the column.

Let $S_{k+1}$ be the matrix :

$S_{k+1} = T_{k+1} + B_{k+1}$

**Proposition 5 :**

For k such $1 \leq k \leq n-1$, we have $(S_{k+1})^{k+1} = \theta$

And $(S_{k+1})^h \neq \theta$ if h is such $1 \leq h \leq k$.

**Proof :**

$(S_{k+1})^2 = (T_{k+1})^2 + (B_{k+1})^2 + B_{k+1} T_{k+1} + T_{k+1} B_{k+1}$

since $(B_{k+1})^2 = \theta$ and $B_{k+1} T_{k+1} = \theta$ and $T_{k+1} B_{k+1} = \theta$ then $(S_{k+1})^2 = (T_{k+1})^2$

Suppose by induction on p that :

$(S_{k+1})^p = (T_{k+1})^p$

then $(S_{k+1})^{p+1} = (T_{k+1})^p (T_{k+1} + B_{k+1}) = (T_{k+1})^{p+1} + (T_{k+1})^p B_{k+1}$

$T_{k+1} B_{k+1} = \theta \Rightarrow (T_{k+1})^p B_{k+1} = \theta$

then

$(S_{k+1})^{p+1} = (T_{k+1})^{p+1}$

By proposition 4, it implies for k such $1 \leq k \leq n-1$, that :

$(S_{k+1})^{k+1} = \theta$ and

$(S_{k+1})^h \neq \theta$ if h is such $1 \leq h \leq k$.

Similarly, for k such $1 \leq k \leq n-2$, we define the matrices $B'_{k+1}$ by :

$(B'_{k+1})_{i,j} \in \{0,1\}$ if $1 \leq i \leq k$ and $k+2 \leq j \leq n$

$(B'_{k+1})_{i,j} = 0$ otherwise.

And we define the matrices $S'_{k+1}$ by:

**Definition** 4 :

$S'_{k+1} = T'_{k+1} + B'_{k+1}$

**Proposition 6:**

For k such $1 \leq k \leq n-1$, we have $(S'_{k+1})^{k+1} = \theta$

and $(S'_{k+1})^h \neq \theta$ if h is such $1 \leq h \leq k$.

**Proof :**

Analogue to proposition 5'proof.

**Proposition 7:**

Let G be a graph of order n , and $U_G$ denote the strictly upper triangular matrix extracted from the adjacency matrix of G, then

$\alpha(G) = n - \min\{h$ such $(U_G)^{h+1} = \theta\}$

$0 \leq h \leq n$
12

**Proof :**

By proposition 1, there exists a split graph S such V(G)=V(S) and E(G)⊂E(S) and

α(G)= α(S)=n-k.

Then $U_G$ is congruent to a matrix $S'_{k+1}$ as defined in definition 4.

Then there exists a permutation matrix P and $P^T$ its transpose matrix such :

$U_G = P^T S'_{k+1} P$

then

$(U_G)^{k+1} = P^T (S'_{k+1})^{k+1} P$

Then :

$(U_G)^{k+1} = P^T \theta P$

And

$(U_G)^h = P^T (S'_{k+1})^h P \neq \theta$ for h such $1 \leq h \leq k$ by proposition 6.

Then k= min{h such $(U_G)^{h+1} = \theta$}

      $0 \leq h \leq n$

And

α(G)= α(S)=n-k=n- min{h such $(U_G)^{h+1} = \theta$}

              $0 \leq h \leq n$

**Proposition 8 :**

Given a graph G=(V,E) on n=│V│ vertices and let $U_G$ denote the strictly upper triangular matrix extracted from the adjacency matrix of G and I the nxn identity matrix, then the independence number α(G) could be computed by the following algorithm :

Input : $U_G$

1 k←0 and T←I

2 T←T$U_G$

3 if T=θ then output(n-k) else k←k+1 goto step 2.

**Example 1 :**

n=5 and G is the complete graph $K_5$

$$U_G = \begin{pmatrix} 0 & 1 & 1 & 1 & 1 \\ 0 & 0 & 1 & 1 & 1 \\ 0 & 0 & 0 & 1 & 1 \\ 0 & 0 & 0 & 0 & 1 \\ 0 & 0 & 0 & 0 & 0 \end{pmatrix}$$

$(U_G)^5 = \theta$ then $\alpha(G) = 5-4 = 1$.

**Example 2 :**

n=5 and G is the following graph :

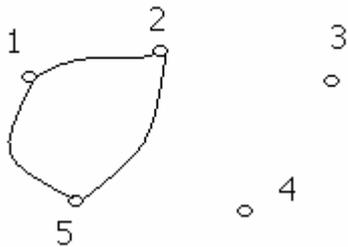

Then

$$U_G = \begin{pmatrix} 0 & 1 & 0 & 0 & 1 \\ 0 & 0 & 0 & 0 & 1 \\ 0 & 0 & 0 & 0 & 0 \\ 0 & 0 & 0 & 0 & 0 \\ 0 & 0 & 0 & 0 & 0 \end{pmatrix}$$

$(U_G)^3 = \theta$ then then $\alpha(G) = 5-2 = 3$.

**Example 3 :**

n=5 and G is the graph 5 $K_1$ (five isolated vertices)

$U_G = \theta$ Then then $\alpha(G) = 5-0 = 5$.